\documentclass[aps,prd,floats,floatfix,showpacs,twocolumn,twoside,preprintnumbers,superscriptaddress,nofootinbib]{revtex4}
\usepackage{graphicx,pstricks}
\usepackage{amsmath,amssymb}
\usepackage{amsfonts}

\begin{document}

\title{BBGKY kinetic approach for an $e^- e^+ \gamma$ plasma created from
the vacuum in a strong laser-generated electric field: 
The one-photon annihilation channel}
\author{D.~B.~Blaschke}
\email{blaschke@ift.uni.wroc.pl}
\affiliation{Institute for Theoretical Physics, University of Wroc{\l}aw, 
50-204 Wroc{\l}aw, Poland}
\affiliation{Bogoliubov Laboratory for Theoretical Physics, Joint Institute 
for Nuclear Research, RU - 141980 Dubna, Russia}
\author{V.~V.~Dmitriev}
\email{smol@sgu.ru}
\affiliation{Department of Physics, Saratov State University, RU - 410026 
Saratov, Russia}
\author{G.~R\"opke}
\email{gerd.roepke@uni-rostock.de}
\affiliation{Institut f\"ur Physik, University of Rostock, D - 18051 Rostock, 
Germany}
\author{S.~A.~Smolyansky}
\email{smol@sgu.ru}
\affiliation{Department of Physics, Saratov State University, RU - 410026 
Saratov, Russia}
\date{\today}

\begin{abstract}
In the present work a closed system of kinetic equations is obtained 
from the truncation of the BBGKY hierarchy for the
description of the vacuum creation of an electron - positron plasma and
secondary photons due to a strong laser field. 
This truncation is performed in the Markovian approximation for the one-photon 
annihilation channel which is accessible due to the presence of the strong 
external field. 
Estimates of the photon production rate are obtained for different
domains of laser field parameters (frequency $\nu$ and field strength $E$). 
A huge quantity of optical photons of the quasiclassical laser
field is necessary to satisfy the conservation laws of the energy
and momentum of the constituents ($e^-, e^+$ and $\gamma$) in this channel.
Since the number of these optical photons corresponds to the order of 
perturbation theory, a vanishingly small photon production rate results 
for the optical region and strongly subcritical fields $E\ll E_{c}$. 
In the $\gamma$-ray region $\nu \lesssim m$ the required number of laser 
photons is small and the production rate of photons from the one-photon
annihilation process becomes accessible to observations for
subcritical fields $E\lesssim E_{c}$. In the infrared region the
photon distribution has a $1/k$ spectrum typical for flicker noise.
\end{abstract}



\maketitle

\section{ Introduction}

The Schwinger effect \cite{SHS} of electron-positron pair (EPP) production
off the vacuum under the action of strong {electromagnetic} fields is
one of the few QED effects that has not yet been tested experimentally.
This is due to the huge electric fields $E\sim E_{c}=m^{2}/e=1.3\cdot
10^{16} $ V/cm necessary to observe the EPP production effect in a constant
external field. Such field strengths are unachievable for static fields.
Therefore, the main attention was devoted to the theoretical and
experimental study of pair creation by time-varying electric fields
generated either in the focal spot of high-power lasers 
\cite{Brezin,PNN,Grib94,Marinov,Bulanov:2004de,Tanji:2008ku} 
or ultraperipheral heavy-ion collisions \cite{b1,b2}. 
Below we will concentrate our attention to EPP creation in strong laser fields. 
The estimations made before in 
Refs.~\cite{Brezin,PNN,Grib94,Marinov,Bunkin,Richards,Troup,Popov} 
showed that pair creation by a single optical laser pulse with $E\ll E_{c}$ 
could hardly be observed. 
More optimistic results have been obtained for planned X-ray free electron 
lasers (XFEL's) \cite{Ringwald,V01,V02} and for counter-propagating laser beams 
in the optical range \cite{Av02,Tar,20,PRL,Ruf:2009zz}.

It is obvious that for subcritical fields $E\ll E_{c}$ the electron -
positron excitations have the character of short-lived quasiparticles which are
not observable after the laser signal ceases. In essence, it is a vacuum
polarization effect. Therefore, the S - matrix methods cannot be used 
\cite{SarRev} and existing estimates \cite{V02} are not reliable. 
An adequate method is provided by the kinetic theory. 
Only on this basis the different experimental manifestations (creation and 
radiation of annihilation photons \cite{Yaresko:2010xe}, generation of harmonics
by focussed laser beams \cite{DiPiazza:2005jc}, birefrigency 
\cite{Heinzl:2006xc} etc. 
\cite{Ruf:2009zz,DiPiazza:2006pr,Dumlu:2010ua,Orthaber:2011cm}) 
find a proper explanation.
Such a kinetic approach has been introduced in \cite{Schmidt:1998vi}, for a 
recent review see \cite{Blaschke:2008wf}. 
However, in these works the photon sector was not yet included consistently
into the quantum kinetic approach. 

In the present work we develop a general kinetic approach where both, the Dirac 
and the photon field are consistently included.
We base our aproach on the BBGKY hierarchy of kinetic equations for the 
electron - positron - photon system generated from the vacuum under the action 
of a time dependent electric field. 
As a first step, we will consider the one-photon annihilation process 
of the quasiparticle EPP,
which in the presence of a strong field is not forbidden \cite{Rit}.
Estimates of the photon production rate for this case constitute the main 
contents of the present work.

The treatment of the two-photon annihilation channel is the next step in 
the development of the quantum kinetic description on the basis of the BBGKY 
hierarchy for the electron - positron - photon system. The detailed
treatment of this important class of processes is delegated to a
subsequent paper. 

Below we will assume an external electric field with the 4 - potential (in
the Hamiltonian gauge) $A^{\mu }(t)=(0,\mathbf{A}(t))$ that is spatially
homogeneous. It is expected that a similar field can be realized
experimentally, e.g., as a standing wave in the small spatial domain of the
focal spot of crossed laser beams. 
The kinetics of quasiparticle EPP creation in vacuum has been investigated in 
detail for the case of a linearly polarized laser field within a 
non-perturbative approach . 
In the case of a rather strong external field 
$\mathbf{A}_{\mathrm{ext}}(t) $ 
some internal field $\mathbf{A}_{\mathrm{int}}(t)$ will be generated too. 
The total acting field will be equal to 
$\mathbf{A}(t)=\mathbf{A}_{\mathrm{int}}(t)+\mathbf{A}_{\mathrm{ext}}(t)$, and 
this field is quasiclassical. 
Fluctuations of the {internal} electromagnetic field  on this 
background can lead to photon excitations that, in principle, can be registered 
outside the active zone of the focal spot.

This article is organized as follows. 
In Sect.~II the system of kinetic equations (KE) in the
quasiparticle representation of the electron - positron
subsystem in a strong time dependent electric field are given. 
The kinetics of the annihilation photons is presented in
Sect.~III and the first equation of the BBGKY chain is obtained. 
In Sect.~IV the
truncation of the BBGKY hierarchy of equations on the level of the
one-photon annihilation process is performed with account of vacuum
polarization effects. 
We intend to clearly state all approximation steps and therefore indicate 
them in the text by ``Approximation 1'' ... ``Approximation 6'' in brackets
at the place of their introduction.
The spectrum of annihilation photons is investigated
here for the large (multiphoton processes) and small (strong fields) values
of the adiabaticity parameter. 
Estimates for the photon radiation rate show that it is very small in the case 
of optical lasers with subcritical fields $E\ll E_{c}$  since a huge number of 
optical photons from the quasiclassical laser field is necessary here to 
overcome the energy gap. 
However, the effect becomes quite
observable in the domain of strong fields $E\lesssim E_{c}$ and high
frequencies (e.g., for the XFEL domain \cite{Ringwald}). Finally, we
summarize the basic results of this work in Sect.~V.

{We use the metric} $g_{\mu \nu }= {\rm diag}(1,-1,-1,-1)$ and
natural units $\hbar =c=1$.

\section{Kinetic equations in the electron - positron sector}

In general, the complete system of equations for a consistent kinetic
description of the electron - positron - photon plasma consists of
\begin{itemize}
\item the KE's for the distribution functions $f_{e,p}(\mathbf{p},t)$ of the
electron and positron quasiparticle components in the presence of the
{strong} quasiclassical total Maxwell field $\mathbf{A}(t)=(0,0,A(t))$;

\item the KE for the distribution function of the photon component, and

\item the Maxwell equation for the quasiclassical internal field $\mathbf{A}%
_{\mathrm{int}}(t)$~.
\end{itemize}
We assume that the electro-neutrality condition holds
\begin{equation}
f_{e}(\mathbf{p},t)=f_{p}(-\mathbf{p},t)=f(\mathbf{p},t)~.  
\label{1e}
\end{equation}
We start from the standard QED Lagrangian 
$\mathcal{L}=\mathcal{L}_{\rm qc}+\mathcal{L}_{I}$ 
taking into account the interaction of the electron-positron Dirac spinor 
fields with a quasiclassical electromagnetic field in $\mathcal{L}_{\rm qc}$ and 
with a quantized one in $\mathcal{L}_{I}$,
\begin{eqnarray}
\mathcal{L}_{\rm qc} &=&\frac{i}{2}\{\overline{\psi }\gamma ^{\mu }D_{\mu }\psi
-(D_{\mu }^{\ast }\overline{\psi })\gamma ^{\mu }\psi \}-m\overline{\psi }
\psi ,  \label{2e} \\
\mathcal{L}_{I} &=&-e\overline{\psi }\gamma ^{\mu }\hat{A}_{\mu }\psi ~,
\label{2f}
\end{eqnarray}
where $D_{\mu }=\partial _{\mu }+ie{A}_{\mu }(t)$. 
Thus, the spatially homogeneous nonstationary {quasiclassical Maxwell} field 
$\mathbf{A}(t)$ is a background to the {quantized} photon field 
$\hat{A}_{\mu}(x)=(0,\hat{\mathbf{A}}(x))$. 
We assume that the intensity of the quantized field is rather weak and its 
influence on the state of the system can be neglected. 
In other words, the electron-positron system plays the role of a photon source. 
The {photon} $in$ - vacuum $|in\rangle$ is defined such that for this state
$\langle \hat{A}_{\mu }(x)\rangle =0~.$

Below we will not consider the backreaction problem because for subcritical
fields $E\ll E_{c}$ the internal field is negligible, {and hence }
$E(t)=E_{\mathrm{ext}}(t)$.

Thus, the idea of the work is the following: With the fermion KE's (see below), 
in principle, the quasiparticle distribution function (\ref{1e}) can be 
determined, taking into account the spin degrees of freedom. 
The separation of the quasiclassical Maxwell field allows to formulate a 
nonperturbative approach to the photon kinetics (Sect.~III). 
In the present work, the photon spectral distribution will be considered mainly 
by including the one-photon annihilation process only (Sect.~IV).


The kinetics of electron - positron vacuum pair creation under the action of
a linearly polarized electric field has been studied in several works, (see,
e.g., \cite{Grib94,Pervushin,Vin01} and the references therein). The
corresponding generalization to the case of an arbitrarily polarized
time-dependent electric field was obtained in \cite{Kiel04,Fil08,PS}. 
In the present work, we will use the oscillator representation
\footnote{In this representation the diagonalization of the Hamiltonian in the
presence of an external field is achieved at once, without the canonical
Bogoliubov transformation procedure.} \cite{Pervushin} which  leads to the 
nonstationary spinor basis \cite{Kiel04,Fil08,PS}
\begin{eqnarray}
u_{1}^{+}(\mathbf{p},t) &=&B(\mathbf{p})[\omega _{+},0,P^{3},P_{-}]~,  \nonumber
\\
u_{2}^{+}(\mathbf{p},t) &=&B(\mathbf{p})[0,\omega _{+},P_{+},-P^{3}]~,
\nonumber \\
v_{1}^{+}(-\mathbf{p},t) &=&B(\mathbf{p})[-P^{3},-P_{-},\omega _{+},0]~,
\nonumber \\
v_{2}^{+}(-\mathbf{p},t) &=&B(\mathbf{p})[-P_{+},P^{3},0,\omega _{+}]~
\label{5e}
\end{eqnarray}
with the usual orthonormalizion conditions \cite{Bogol}
\begin{gather}
{u}_{\alpha }^{+}(\mathbf{p},t){v_{\beta }}(-\mathbf{p},t)=0~,  
\nonumber \\
u_{\alpha }^{+}(\mathbf{p},t)u_{\beta }(\mathbf{p},t)=
v_{\alpha }^{+}(-\mathbf{p},t)v_{\beta }(-\mathbf{p},t)=
\delta _{\alpha \beta }~,  
\nonumber \\
{\bar{u}}_{\alpha }(\mathbf{p},t)u_{\beta }(\mathbf{p},t)=
\frac{m}{\omega (\mathbf{p},t)}{\delta }_{\alpha \beta }~,  
\nonumber \\
{\bar{v}}_{\alpha }(\mathbf{p},t)v_{\beta }(\mathbf{p},t)=
-\frac{m}{\omega (\mathbf{p},t)}{\delta }_{\alpha \beta }~ , 
\label{6e}
\end{gather}
where $\omega (\mathbf{p},t)=\sqrt{m^{2}+\mathbf{P}^{2}}$ , 
$\mathbf{P}=\mathbf{p}-e\mathbf{A}(t)$, $P_{\pm }=P^{1}\pm iP^{2}$, 
$\omega _{+}=\omega +m$ and $B(\mathbf{p})=[2\omega \omega _{+}]^{-1/2}$. 
The Hamiltonian has the diagonal form in this quasiparticle representation
\begin{eqnarray}
H_{f}(t)=\sum_{\alpha }\int d^{3}p&&\omega (\mathbf{p},t)
\big[a_{\alpha}^{+}(\mathbf{p},t)a_{\alpha }(\mathbf{p},t)
\nonumber\\
&&-b_{\alpha }(-\mathbf{p},t)b_{\alpha }^{+}(-\mathbf{p},t)\big]\ .  
\label{8e}
\end{eqnarray}

Finally, the Dirac equation in the presence of the external quasiparticle
field $\mathbf{A}_{\mathrm{ext}}(t)$ generates the Heisenberg - like
equations of motion for the time-dependent creation and annihilation
operators
\begin{eqnarray}
\dot{a}(\mathbf{p},t) &=&-U_{(1)}(\mathbf{p},t)a(\mathbf{p},t)
                         -U_{(2)}(\mathbf{p},t)b^{+}(-\mathbf{p},t)\nonumber\\
			&&-i\omega (\mathbf{p},t)a(\mathbf{p},t),
\nonumber \\
\dot{b}(-\mathbf{p},t) &=&b(-\mathbf{p},t)U_{(1)}(\mathbf{p},t)
			+a^{+}(\mathbf{p},t)U_{(2)}(\mathbf{p},t)\nonumber\\
			&&-i\omega (\mathbf{p},t)b(-\mathbf{p},t)~,
\label{9e}
\end{eqnarray}
with the matrices in the representation (\ref{5e}) being 
\begin{eqnarray}
U_{(1)}(\mathbf{p},t) &=&
i\omega C(\mathbf{p})[\mathbf{P}\mathbf{E}]\mathbf{{\boldsymbol{\sigma }}}=
i\mathbf{U\sigma },  
\label{10e} \\
U_{(2)}(\mathbf{p},t) &=&\mathbf{q}\mathbf{\boldsymbol{\sigma }}=
C(\mathbf{p})[\mathbf{P}(\mathbf{P}\mathbf{E})
-\mathbf{E}\omega \omega _{+}]\mathbf{\boldsymbol{\sigma }}~.  
\label{10.5e}
\end{eqnarray}
Here, $\mathbf{E}(t)=-\dot{\mathbf{A}}(t)$ is the strength of the external
electric field for $|\mathbf{A}_{\mathrm{int}}|\ll |\mathbf{A}_{\mathrm{ext}}|$
and $C(\mathbf{p})=e/(2\omega^{2}\omega _{+})$. 
The matrices (\ref{10e}) and (\ref{10.5e}) describe the
different vacuum effects in the presence of an external electric field
(polarization, spin rotation, EPP creation).

The KE's for the electron - positron component of the plasma follow from the
equations of motion (\ref{9e}) and the definitions of the electron and
positron distribution functions in the instantaneous representation 
\cite{Grib94}.
\begin{eqnarray}
f_{\alpha \beta }(\mathbf{p},t) &=&
\langle a_{\beta }^{+}(\mathbf{p},t)a_{\alpha }(\mathbf{p},t)\rangle ~,  
\nonumber \\
{f}_{\alpha \beta }^{c}(\mathbf{p},t) &=&
\langle b_{\beta }(-\mathbf{p},t)b_{\alpha }^{+}(-\mathbf{p},t)\rangle ~,  
\label{11e}
\end{eqnarray}
and also two additional functions
\begin{eqnarray}
f_{\alpha \beta }^{(+)}(\mathbf{p},t) &=&
\langle a_{\beta }^{+}(\mathbf{p},t)b_{\alpha }^{+}(-\mathbf{p},t)\rangle ,  
\nonumber \\
f_{\alpha \beta }^{(-)}(\mathbf{p},t) &=&
\langle b_{\beta }(-\mathbf{p},t)a_{\alpha }(\mathbf{p},t)\rangle ~,  
\label{12e}
\end{eqnarray}
describing vacuum polarization. 
The KE system in matrix notation is then
\cite{Kiel04,Fil08,PS}
\begin{eqnarray}
\dot{f} &=&[f,U_{(1)}]-\left( U_{(2)}f^{(+)}+f^{(-)}U_{(2)}\right) ~, 
\nonumber\\
\dot{f}^{c} &=&[f^{c},U_{(1)}]+\left( f^{(+)}U_{(2)}+U_{(2)}f^{(-)}\right) ~,
\nonumber \\
\dot{f}^{(+)} &=&[{f}^{(+)},U_{(1)}]+\left( U_{(2)}f-f^{c}U_{(2)}\right)
+2i\omega f^{(+)}~,  
\nonumber \\
\dot{f}^{(-)} &=&[{f}^{(-)},U_{(1)}]+\left( fU_{(2)}-U_{(2)}f^{c}\right)
-2i\omega f^{(-)}~.
\nonumber\\ 
 \label{13e}
\end{eqnarray}

If the standard decomposition in the basis of Pauli matrices is used ($k=1,2,3$)
\begin{equation}
f = f_0+f_k \sigma_k~,  
\label{14e}
\end{equation}
where $f_0 = \frac{1}{2}{\rm Tr} f$ , $f_k=\frac{1}{2}{\rm Tr} f\sigma_k$, 
the KE's (\ref{13e}) can be rewritten in the spin representation, where the 
first set of equations concerns the spinless distributions
\begin{eqnarray}
\dot{f}_{0}&=&-2\mathbf{q}\mathbf{u}~,  \nonumber \\
\dot{f}_{0}^{c}&=&2\mathbf{q}\mathbf{u}~,  \nonumber \\
\dot{u}_{0}&=&2\omega v_{0}+(\mathbf{f}-\mathbf{f}^{c})\mathbf{q}~, \nonumber\\
\dot{v}_{0}&=&-2\omega u_{0}~,  
\label{15e}
\end{eqnarray}
and the KE's for the spin distribution functions are collected in the second set
\begin{eqnarray}
\dot{f}_{k}&=& -2u_{0}q_{k}-2[\mathbf{f}\mathbf{U}]_{k}
+2[\mathbf{v}\mathbf{q}\,]_{k}~,  \nonumber \\
\dot{f}_{k}^{c}&=& 2u_{0}q_{k}-2[\mathbf{f}^{c}\mathbf{U}]_{k}
-2[\mathbf{v}\mathbf{q}\,]_{k}~,  \nonumber \\
\dot{u}_{k}&=& 2\omega v_{k}-2[\mathbf{u}\mathbf{U}]_{k}
+({f}_{0}-{f}_{0}^{c})q_{k}~,  \nonumber \\
\dot{v}_{k}&=& -2\omega u_{k}-2[\mathbf{v}\mathbf{U}]_{k} 
+[(\mathbf{f}+\mathbf{f}^{c})\mathbf{q}\,]_{k}~.  
\label{15f}
\end{eqnarray}
Some simple applications of this system of KE's can be found in 
\cite{Kiel04,Fil08,PS}.

In the simplest case of the linearly polarized photon field 
$\hat{\mathbf{A}}(t)=(0,0,A(t))$ 
the system of equations (\ref{15e}) is transformed into a coupled system of 
three ordinary first-order differential equations 
\cite{Blaschke:2008wf,Schmidt:1998vi}
\begin{eqnarray}
\dot{f} &=&\frac{1}{2}\lambda u~,  \nonumber \\
\dot{u} &=&\lambda (1-2f)-2\omega v~,  \nonumber \\
\dot{v} &=&2\omega u~,  \label{KE-8}
\end{eqnarray}
where $\lambda (\mathbf{p},t)=eE(t)\varepsilon _{\perp }/\omega ^{2}$, 
$\varepsilon _{\perp }(\mathbf{p})=\sqrt{m^{2}+\mathbf{p}_{\perp }^{2}}$ 
and $\mathbf{p}_{\perp }=(p^{(1)},p^{(2)},0)$. 
The system (\ref{KE-8}) corresponds to a KE of the non-Markovian type
\begin{eqnarray}
f(\mathbf{p},t)=\frac{1}{2}\lambda (\mathbf{p},t)\int_{t_{0}}^{t}
&&dt^{\prime}\lambda (\mathbf{p},t^{\prime })[1-2f(\mathbf{p},t^{\prime })]
\nonumber\\
&&\cos \left[2\int_{t^{\prime }}^{t}d\tau \omega (\mathbf{p},\tau )\right] ~.
\label{KE-9}
\end{eqnarray}
This KE and its representation (\ref{KE-8}) have been used in numerous
applications, see \cite{Blaschke:2008wf}.

The further development of the kinetic theory approach to vacuum
particle creation is progressing in search of different exact 
and approximate solutions of the problem (see, e.g., 
\cite{Hebenstreit:2008ae,Hebenstreit:2009km,Fedotov,Dumlu:2011rr})
and also in extending the possibilities of the formalism 
\cite{Hebenstreit:2010vz}.

\section{Photon sector}

\subsection{The basic equations}

An external electric field that generates an instability of the vacuum with
respect to electron - positron pair creation is accompanied by the
appearance of internal currents and electromagnetic fields (back reaction
problem). Quantum fluctuations of this internal field are interpreted as
photon excitations, which can leave the active zone (focal spot) and can be
registered experimentally. Below the KE for the description of the photon
component will be obtained and investigated for some simple situations.

For the construction of the photon kinetics it is necessary to develop the
corresponding generalization of the quasiparticle formalism developed in
Sect.~II using it as a nonperturbative basis.

Let us introduce at first the interaction with the quantized electromagnetic
field $\hat{A}_{\mu }(x)$ in the fermion sector of the theory by means of
the substitution $H_{f}\rightarrow H_{f}+H_{\rm int}$ in the Heisenberg-like
equation of motion (\ref{9e})
\begin{widetext}
\begin{eqnarray}
\dot{a}(\mathbf{p},t)+U_{(1)}(\mathbf{p},t)a(\mathbf{p},t)
		+U_{(2)}(\mathbf{p},t)b^{+}(-\mathbf{p},t) 
&=&-i[a(\mathbf{p},t),H_{f}+H_{\rm int}]  
\nonumber \\
\dot{b}(-\mathbf{p},t)-b(-\mathbf{p},t)U_{(1)}(\mathbf{p},t)
		+a^{+}(\mathbf{p},t)U_{(2)}(\mathbf{p},t) 
&=&-i[b(-\mathbf{p},t),H_{f}+H_{\rm int}]~,  
\label{16e}
\end{eqnarray}
\end{widetext}
where $H_{f}$ is the Hamiltonian (\ref{8e}) of the fermion field in the
quasiparticle representation and $H_{\mathrm{int}}$ is the usual Hamiltonian
of the interaction with the quantized field
\begin{equation}
H_{\mathrm{int}}(t)=e\int d^{3}x
	:\overline{\psi }(x)\gamma ^{\mu }\hat{A}_{\mu }(x)\psi (x):~.  
\label{17e}
\end{equation}
The time dependence of $H_{\rm int}(t)$ is due to the nonstationarity of the
system that is reflected in the characteristic oscillator representation of
the decomposition of the field operators $\psi (x),\,\psi ^{+}(x)$ in the
nonstationary basis (\ref{5e}), (\ref{6e}). 
The same source (external field) generates the nonstationarity 
of the quantized electromagnetic field. 
However, this does not induce an alteration of the mass shell of the photon 
field, $k^\mu k_\mu=0$ (in contrast to the electron - positron field, 
$\omega (\mathbf{k},t)$), and then the standard decomposition is valid,
\begin{equation}
\hat{A}_{\mu }(x)=(2\pi )^{-3/2}\int \frac{d^{3}k}{\sqrt{2k}}
		A_{\mu }(\mathbf{k},t)e^{-i\mathbf{k}\mathbf{x}}~,  
\label{18e}
\end{equation}
where 
$A_{\mu }(\mathbf{k},t)=A_{\mu }^{(+)}(\mathbf{k},t)
+A_{\mu }^{(-)}(-\mathbf{k},t)$ 
with the condition 
$\{A_{\mu }^{(\pm )}(\mathbf{k},t)\}^{+}=A_{\mu }^{(\mp )}(-\mathbf{k},t)$ 
and the standard commutation relations
\begin{equation}
\lbrack A_{\mu }^{(-)}(\mathbf{k},t),A_{\mu }^{(+)}(\mathbf{k^{\prime }},t)]
=-g_{\mu \nu }\delta (\mathbf{k}-\mathbf{k^{\prime }})~.  
\label{19e}
\end{equation}

The Hamiltonian of the interaction (\ref{17e}) in the oscillator
representation has the form
\begin{eqnarray}
H_{\rm int}(t) &=&e(2\pi )^{-3/2}\sum_{\alpha \beta }\int d^{3}p_{1}d^{3}p_{2}
\frac{d^{3}k}{\sqrt{2k}}\;
\nonumber\\
&&\delta (\mathbf{p_{1}}-\mathbf{p_{2}}+\mathbf{k})
\nonumber \\
&:&\bigl\{
[\bar{u}v]_{\beta \alpha }^{r}(\mathbf{p}_{1},\mathbf{p}_{2},\mathbf{k};t)
a_{\alpha }^{+}(\mathbf{p}_{1},t)a_{\beta }(\mathbf{p}_{2},t)
\nonumber \\
&&+
[\bar{u}v]_{\beta \alpha }^{r}(\mathbf{p}_{1},\mathbf{p}_{2},\mathbf{k};t)
a_{\alpha }^{+}(\mathbf{p}_{1},t)b_{\beta }^{+}(-\mathbf{p}_{2},t)  
\nonumber \\
&&+
[\bar{v}u]_{\beta \alpha }^{r}(\mathbf{p}_{1},\mathbf{p}_{2},\mathbf{k};t)
b_{\alpha }(-\mathbf{p}_{1},t)a_{\beta }(\mathbf{p}_{2},t)  
\nonumber \\
&&+
[\bar{v}v]_{\beta \alpha }^{r}(\mathbf{p}_{1},\mathbf{p}_{2},\mathbf{k};t)
b_{\alpha }(-\mathbf{p}_{1},t)b_{\beta }^{+}(-\mathbf{p}_{2},t)\bigr\}
\nonumber\\
&&A_{r}(\mathbf{k},t):~~.  
\label{20e}
\end{eqnarray}
Here and below the vectors $\mathbf{p}_{1},\mathbf{p}_{2},\ldots $ are used
for notation of the canonical momenta of fermions and 
$\mathbf{k}_{1},\mathbf{k}_{2},\ldots $ 
correspond to the momenta of photons. 
The spinor convolutions are
\begin{equation}
[\bar{\xi}\eta ]_{\beta \alpha}^{r}(\mathbf{p}_{1},\mathbf{p}_{2},\mathbf{k};t)
=\bar{\xi}_{\alpha }(\mathbf{p}_{1},t)\gamma ^{\mu }
\eta_{\beta }(\mathbf{p}_{2},t)e_{\mu }^{r}(\mathbf{k})~,  
\label{21e}
\end{equation}
where the spinors 
$\xi _{\alpha }(\mathbf{p},t)$ and $\eta _{\alpha }(\mathbf{p},t)$ 
are taken from the set 
$\{u_{\alpha }(\mathbf{p},t),v_{\alpha}(-\mathbf{p},t)\}$ (\ref{5e}), and 
$\mathbf{e}^{\;1},\mathbf{e}^{\;2}$ 
are the photon polarization unit vectors, while 
$\mathbf{e}^{\;3}=\mathbf{k}/k$
and $e_{\mu }^{0}=\delta _{\mu }^{0}$. 
The relationship \cite{Bogol}
\begin{equation}
A_{\mu }^{(\pm )}(\mathbf{k},t)=
\sum_{r=0}^{3}e_{\mu }^{r}A_{r}^{(\pm )}(\mathbf{k},t)  
\label{22e}
\end{equation}
has been used here.

The Hamiltonian of the free photon field is given by
\begin{equation}
H_{ph}(t)=\sum_{r=1,2}\int d^3 k\; kA^{(+)}_{r}(\mathbf{k},t)A^{(-)}_{r}(%
\mathbf{k},t).  \label{23e}
\end{equation}

The system of Heisenberg-like equations of motion with account of the photon
subsystem (\ref{16e}) can be written now in explicit form. 
For example,
\begin{eqnarray}
\dot{a}(\mathbf{p},t) &=&-i\omega (\mathbf{p},t)a(\mathbf{p},t)
-U_{(1)}(\mathbf{p},t)a(\mathbf{p},t)
\nonumber\\
&&-U_{(2)}(\mathbf{p},t)b^{+}(-\mathbf{p},t)
\nonumber \\
&&+ie(2\pi )^{-3/2}\int d^{3}p_{1}\frac{d^{3}k}{\sqrt{2k}}\;
\delta (\mathbf{p}-\mathbf{p}_{1}+\mathbf{k})
\nonumber\\
&&\Bigl\{
a(\mathbf{p}_{1},t)[\bar{u}u]^{r}(\mathbf{p},\mathbf{p}_{1},\mathbf{k};t)  
\nonumber \\
&&+b(-\mathbf{p}_{1},t)[\bar{u}v]^{r}(\mathbf{p},\mathbf{p}_{1},\mathbf{k};t)
\Bigr\}A_{r}(\mathbf{k},t)~,  
\label{24e_1}
\\
\dot{b}(-\mathbf{p},t) &=&
	-i\omega (\mathbf{p},t)b(-\mathbf{p},t)
	+b(-\mathbf{p},t)U_{(1)}(\mathbf{p},t)
\nonumber\\
&&	+a^{+}(\mathbf{p},t)U_{(2)}(\mathbf{p},t)  
\nonumber \\
&&-ie(2\pi )^{-3/2}\int d^{3}p_{1}\frac{d^{3}k}{\sqrt{2k}}\;
\delta (\mathbf{p}_{1}-\mathbf{p}+\mathbf{k})
\nonumber\\
&&\Bigl\{
[\bar{u}v]^{r}(\mathbf{p}_{1},\mathbf{p},\mathbf{k};t)a^{+}(\mathbf{p}_{1},t)  
\nonumber \\
&&+[\bar{v}v]^{r}(\mathbf{p}_{1},\mathbf{p},\mathbf{k};t)b(-\mathbf{p}_{1},t)
\Bigr\}A_{r}(\mathbf{k},t)~,  
\label{24e_2}
\\
iA_{r}^{(\pm )}(\mathbf{k},t) &=&
	\mp kA_{r}^{(\pm )}(\mathbf{k},t)
\nonumber\\
&&	\mp e(2\pi)^{-3/2}\frac{1}{\sqrt{2k}}
	\int d^{3}p_{1}d^{3}p_{2}\;
	\delta (\mathbf{p}_{1}-\mathbf{p}_{2}\mp \mathbf{k})  
\nonumber \\
&&\Bigl\{
a^{+}(\mathbf{p}_{1},t)
[\bar{u}u]^{r}(\mathbf{p}_{1},\mathbf{p}_{2},\mathbf{k};t)
a(\mathbf{p}_{2},t)
\nonumber\\
&&+a^{+}(\mathbf{p}_{1},t)
[\bar{u}v]^{r}(\mathbf{p}_{1},\mathbf{p}_{2},\mathbf{k};t)
b^{+}(-\mathbf{p}_{2},t)  
\nonumber\\
&&+b(-\mathbf{p}_{1},t)
[\bar{v}u]^{r}(\mathbf{p}_{1},\mathbf{p}_{2},\mathbf{k};t)a(\mathbf{p}_{2},t)
\nonumber\\
&&
+b(-\mathbf{p}_{1},t)[\bar{v}v]^{r}(\mathbf{p}_{1},\mathbf{p}_{2},\mathbf{k};t)
b^{+}(-\mathbf{p}_{2},t)\Bigr\}~. \nonumber\\ 
\label{24e_3}
\end{eqnarray}
This is the exact system of equations with nonperturbative
account of the external electric field.

\subsection{The first equation of the BBGKY hierarchy}

We start from the single - time, two - point photon correlation function in
momentum space
\begin{equation}
F_{rr^{\prime}}(\mathbf{k},\mathbf{k}^{\prime},t)= 
\langle A^{(+)}_{r}(\mathbf{k},t) 
	A^{(-)}_{r^{\prime}}(\mathbf{k}^{\prime},t)\rangle~.
\label{25e}
\end{equation}
In the spatially homogeneous case it is diagonal in the momentum $\mathbf{k}$ 
and polarization $r$ with the photon distribution function $F_r(\mathbf{k},t)$ 
as matrix element,
\begin{equation}
\langle A^{(+)}_{r}(\mathbf{k},t) 
	A^{(-)}_{r^{\prime}}(\mathbf{k}^{\prime},t)\rangle 
= \delta_{rr^{\prime}}\delta(\mathbf{k}
-\mathbf{k^{\prime}})F_r (\mathbf{k},t)~.  
\label{25_2}
\end{equation}
The diagonalization in the polarization indices is an approximation here
(Approximation 1).

Then from the commutation relation (\ref{19e}) it follows
\begin{equation}
\langle A^{(-)}_{r}(\mathbf{k},t)
	A^{(+)}_{r^{\prime}}(\mathbf{k}^{\prime},t)\rangle
= \delta_{rr^{\prime}}\delta(\mathbf{k}
  -\mathbf{k^{\prime}})\{1+F_r (\mathbf{k},t)\}.  
\label{25_3}
\end{equation}

Let us write now the first equation of the BBGKY hierarchy for the
correlation function (\ref{25e}) using the photon equation of motion 
(\ref{24e_3})
\begin{widetext}
\begin{eqnarray}
\dot{F}_{rr^{\prime}}(\mathbf{k},\mathbf{k}^{\prime},t) &=&
ie(2\pi)^{-3/2}\sum_{\alpha,\beta}\int d^3 p_1 d^3 p_2 
\Bigg\{-\frac{1}{\sqrt{2k}} \delta(\mathbf{p}_1 -\mathbf{p}_2 -\mathbf{k})  
\bigg[[\bar{u}u]^{r}_{\beta\alpha}(\mathbf{p}_1 ,\mathbf{p}_2 ,\mathbf{k};t) 
\langle a^{+}_{\alpha}(\mathbf{p}_{1},t)a_{\beta}(\mathbf{p}_2,t)
A_{r^{\prime}} ^{(-)} (\mathbf{k}^{\prime},t)\rangle  
\nonumber \\
&&+ [\bar{u}v]^{r}_{\beta\alpha}(\mathbf{p}_1 ,\mathbf{p}_2 ,\mathbf{k};t)
\langle a^{+}_{\alpha}(\mathbf{p}_{1},t)b_{\beta}^+ (-\mathbf{p}_2,t)
A_{r^{\prime}} ^{(-)} (\mathbf{k}^{\prime},t)\rangle  
+ [\bar{v}u]^{r}_{\beta\alpha}(\mathbf{p}_1 ,\mathbf{p}_2 ,\mathbf{k};t)
\langle b_{\alpha}(-\mathbf{p}_{1},t)a_{\beta}(\mathbf{p}_2 ,t)
A_{r^{\prime}}^{(-)} (\mathbf{k}^{\prime},t)\rangle  
\nonumber \\
&& + [\bar{v}v]^{r}_{\beta\alpha}(\mathbf{p}_1 ,\mathbf{p}_2 ,\mathbf{k};t)
\langle b_{\alpha}(-\mathbf{p}_{1},t)b_{\beta}^+ (-\mathbf{p}_2,t)
A_{r^{\prime}} ^{(-)} (\mathbf{k}^{\prime},t)\rangle \bigg]  
+ \frac{1}{\sqrt{2k^{\prime}}} 
\delta(\mathbf{p}_1 -\mathbf{p}_2 +\mathbf{k}^{\prime})  
\nonumber \\
&& \bigg[
[\bar{u}u]^{r^{\prime}}_{\beta\alpha} 
(\mathbf{p}_1 ,\mathbf{p}_2,\mathbf{k}^{\prime};t) 
\langle a^{+}_{\alpha}(\mathbf{p}{1},t)a_{\beta}(\mathbf{p}_2 ,t) 
A_{r} ^{(+)} (\mathbf{k},t)\rangle  
+ [\bar{u}v]^{r^{\prime}}_{\beta\alpha} 
(\mathbf{p}_1 ,\mathbf{p}_2 ,\mathbf{k}^{\prime};t) 
\langle a^{+}_{\alpha}(\mathbf{p}_{1},t)b_{\beta}^+ (-\mathbf{p}_2,t) 
A_{r} ^{(+)} (\mathbf{k},t)\rangle  
\nonumber \\
&&+ [\bar{v}u]^{r^{\prime}}_{\beta\alpha} 
(\mathbf{p}_1 ,\mathbf{p}_2 ,\mathbf{k}^{\prime};t) 
\langle b_{\alpha}(-\mathbf{p}_{1},t)a_{\beta} (\mathbf{p}_2,t) 
A_{r} ^{(+)} (\mathbf{k},t)\rangle  
+ [\bar{v}v]^{r^{\prime}}_{\beta\alpha} 
(\mathbf{p}_1 ,\mathbf{p}_2 ,\mathbf{k}^{\prime};t) 
\langle b_{\alpha}(-\mathbf{p}_{1},t)b_{\beta}^+ (-\mathbf{p}_2 ,t) 
A_{r} ^{(+)} (\mathbf{k},t)\rangle \bigg]\Bigg\}  
\nonumber \\
&&+i({k}-{k}^{\prime}){F}_{rr^{\prime}}(\mathbf{k},\mathbf{k}^{\prime},t) ~.
\label{26e}
\end{eqnarray}
\end{widetext}
The last term on the r.h.s. of Eq.~(\ref{26e}) describes the quantum beating
of two-photon states and can be omitted in the approximation (\ref{25_2}).

Thus, the kinetics of the photon states is defined by the different forced
processes of either one-photon scattering of electron and positron
(considered as quasiparticles) or their creation and annihilation. Some
processes forbidden in absence of an external field become possible here:
e.g., in the lowest order of the perturbation theory it is the one-photon
annihilation, the simultaneous creation of an electron - positron pair and a
photon \cite{Rit}. Apparently, the latter processes are strongly suppressed
in the region of subcritical fields. In highest orders of perturbation
theory the number of this kind of forbidden processes increases abruptly. In
the general case, the necessity of taking into account such kind of
processes strongly complicates the problem. 

According to the approach given in Sect.~II, the interaction of
the photon subsystem with the electron-positron one can be taken into
account in the framework of the standard perturbation theory with the
fine-structure constant $\alpha =e^{2}/(4\pi) $ as an expansion parameter.
This justifies the truncation of the BBGKY 
hierarchy for the correlators occurring in Eq.~(\ref{26e}).

\section{One-photon annihilation channel}

\subsection{Annihilation channel}

Below we shall restrict the discussion to the annihilation channel only.
Keeping the relevant terms in Eq.~(\ref{26e}) ('incoming - outgoing' terms,
the second line in the first square bracket in r.h.s. of Eq.~(\ref{26e}) and
the third line in the second bracket), we obtain

\begin{eqnarray}
\dot{F}_{rr^{\prime}}(\mathbf{k},\mathbf{k}^{\prime},t)
&=&ie(2\pi)^{-3/2}\int d^3 p_1 d^3 p_2 
\nonumber\\
&&\Bigl\{\frac{1}{\sqrt{2k}} \delta(\mathbf{p}_1 -\mathbf{p}_2 -\mathbf{k})  
[\bar{u}v]^{r}_{\beta\alpha}(\mathbf{p}_1 ,\mathbf{p}_2 ,\mathbf{k};t) 
\nonumber\\
&&\langle b^{+}_{\beta}(-\mathbf{p}_{2},t)a^+_{\alpha}(\mathbf{p}_1,t)
A_{r^{\prime}}^{(-)}(\mathbf{k}^{\prime},t)\rangle  
\nonumber
 \\
&+& \frac{1}{\sqrt{2k^{\prime}}} 
\delta(\mathbf{p}_1 -\mathbf{p}_2 +\mathbf{k}^{\prime}) 
[\bar{v}u]^{r^\prime}_{\beta\alpha} 
(\mathbf{p}_1,\mathbf{p}_2,\mathbf{k}^{\prime};t) 
\nonumber\\
&&\langle b_\alpha(-\mathbf{p}_1,t)a_{\beta}(\mathbf{p}_2,t) 
A_{r}^{(+)}(\mathbf{k},t)\rangle \Bigr\}~.  
\label{27e}
\end{eqnarray}

For obtaining a closed photon KE it is necessary to perform some truncation
procedure for the correlators occurring in this equation. The simplest
truncation of the type
\begin{eqnarray}
&&\langle b_{\alpha}(-\mathbf{p}_1,t)a_\beta(\mathbf{p}_2,t) 
A_{r} ^{(\pm)} (\mathbf{k},t)\rangle \simeq 
\nonumber\\
&&\hspace{1cm}
\langle b_{\alpha}(-\mathbf{p}_{1},t)a_{\beta}(\mathbf{p}_2,t)\rangle 
\langle A_{r} ^{(\pm)} (\mathbf{k}^{\prime},t)\rangle
=0  
\label{28e}
\end{eqnarray}
is not effective due to the definition of the photon vacuum $\langle A_r
^{(\pm)}(\mathbf{k},t)\rangle=0$ (Sect.~II).

The equations of the second order for the correlators from Eq.~(\ref{27e})
can be obtained easily with help of {the equations of motion~}
(\ref{24e_1})-(\ref{24e_3}), 
e.g.,
\begin{widetext}
\begin{eqnarray}
\biggl\{\frac{\partial}{\partial t}&+& 
i[\omega(\mathbf{p}_1,t)+\omega(\mathbf{p}_2,t)-k]\biggr\} 
\langle b_{\alpha}(-\mathbf{p}_{1},t)a_{\beta}(\mathbf{p}_2 ,t) 
A_{r} ^{(+)} (\mathbf{k},t)\rangle  
=-ie(2\pi)^{-3/2}\int d^3 p^{\prime} \frac{d^3 k^{\prime}}{\sqrt{2k^{\prime}}}
\Bigl\{ \delta(\mathbf{p}^{\prime}-\mathbf{p}_1 +\mathbf{k}^{\prime}) 
\nonumber \\
&& \left[ 
[\bar{u}v]^{r^\prime}_{\alpha\beta^\prime} 
(\mathbf{p}^\prime,\mathbf{p}_1,\mathbf{k}^{\prime};t) 
\langle a^{+}_{\beta^{\prime}}(\mathbf{p}^{\prime},t) 
a_{\beta}(\mathbf{p}_2 ,t)A_{r^{\prime}} (\mathbf{k}^{\prime},t) 
A_{r}^{(+)} (\mathbf{k},t)\rangle  \right.
\nonumber \\
&+&\left. [\bar{v}v]^{r^{\prime}}_{\alpha\beta^{\prime}} 
(\mathbf{p}^{\prime},\mathbf{p}_1 ,\mathbf{k}^{\prime};t) 
\langle b_{\beta^{\prime}}(-\mathbf{p}^{\prime},t) 
a_{\beta}(\mathbf{p}_2 ,t) A_{r^{\prime}} (\mathbf{k}^{\prime},t) 
A_{r} ^{(+)} (\mathbf{k},t)\rangle \right]  
\nonumber \\
&-&\delta(\mathbf{p}_2 -\mathbf{p}^{\prime}+\mathbf{k}^{\prime}) \cdot \left[
[\bar{u}u]^{r^{\prime}}_{\beta^{\prime}\beta} (\mathbf{p}_2 ,\mathbf{p}%
^{\prime},\mathbf{k}^{\prime};t) \langle b_{\alpha}(-\mathbf{p}%
_1,t)a_{\beta^{\prime}}(\mathbf{p}^{\prime},t) A_{r^{\prime}} (\mathbf{k}%
^{\prime},t) A_{r}^{(+)} (\mathbf{k},t)\rangle \right.  \nonumber \\
&+& \left.[\bar{u}v]^{r^{\prime}}_{\beta^{\prime}\beta} (\mathbf{p}_2 ,%
\mathbf{p}^{\prime},\mathbf{k}^{\prime};t) \langle b_{\alpha}(-\mathbf{p}%
_1,t) b_{\beta^{\prime}}^+ (-\mathbf{p}^{\prime},t) A_{r^{\prime}} (\mathbf{k%
}^{\prime},t) A_{r} ^{(+)} (\mathbf{k},t)\rangle \right] \Bigr\}  \nonumber \\
&+& S^r_{\alpha\beta}(\mathbf{p}_1 ,\mathbf{p}_2 ,\mathbf{k};t)
+U^r_{\alpha\beta}(\mathbf{p}_1 ,\mathbf{p}_2 ,\mathbf{k};t)~.  \label{29e}
\end{eqnarray}
\end{widetext}
On the r.h.s. of this equation there is a set of the terms originating from
vacuum polarization effects in the presence of the quantized electromagnetic
field (Sect.~II), which are absent in the standard QED without a strong
field. The group of terms
\begin{widetext}
\begin{eqnarray}
U^r _{\alpha\beta}(\mathbf{p}_1 ,\mathbf{p}_2 ,\mathbf{k};t) &=&
U^{(1)}_{\beta^{\prime}\alpha}(\mathbf{p}_1 ,t) 
\langle b_{\beta^{\prime}}(-\mathbf{p}_{1},t) a_{\beta}(\mathbf{p}_2 ,t) 
A_{r} ^{(+)} (\mathbf{k},t)\rangle  
- U^{(1)} _{\beta\beta^{\prime}}(\mathbf{p}_2 ,t) 
\langle b_{\alpha}(-\mathbf{p}_{1},t) a_{\beta^{\prime}}(\mathbf{p}_2 ,t) 
A_{r} ^{(+)} (\mathbf{k},t)\rangle  
\nonumber \\
&+& U^{(2)} _{\beta^{\prime}\alpha}(\mathbf{p}_1 ,t) \langle
a_{\beta^{\prime}}^+(\mathbf{p}_{1},t) a_{\beta}(\mathbf{p}_2 ,t) 
A_{r}^{(+)} (\mathbf{k},t)\rangle  
- U^{(2)} _{\beta\beta^{\prime}}(\mathbf{p}_2 ,t) 
\langle b_{\alpha}(-\mathbf{p}_{1},t) b_{\beta^{\prime}}(-\mathbf{p}_2 ,t) 
A_{r} ^{(+)} (\mathbf{k},t) \rangle 
\label{30e}
\end{eqnarray}
\end{widetext}
can be omitted because of the approximation (\ref{28e}) in order to close
the chain of equations at this minimal level.

The other set of terms leads to contributions of the vacuum polarization 
effects in the one-photon annihilation channel \cite{Smolyansky:2010as}

\begin{widetext}
\begin{eqnarray}
S^r_{\alpha\beta}(\mathbf{p}_1 ,\mathbf{p}_2 ,\mathbf{k};t) &=& 
- ie(2\pi)^{-3/2}\frac{1}{\sqrt{2k}}\int d^3 p_1^{\prime}d^3 p^{\prime}_2
\delta(\mathbf{p}^{\prime}_1 -\mathbf{p}^{\prime}_2 -\mathbf{k})  
\nonumber \\
&& \left\{ [\bar{u}u]^{r}_{\alpha^{\prime}\beta^{\prime}} 
	(\mathbf{p}^{\prime}_1 ,\mathbf{p}^{\prime}_2 ,\mathbf{k};t) 
	\langle b_{\alpha}(-\mathbf{p}_1,t) a_{\beta}(\mathbf{p}_2 ,t) 
		a^+_{\alpha^{\prime}}(\mathbf{p}^{\prime}_1,t) 
		a_{\beta^{\prime}}(\mathbf{p}^{\prime}_2 ,t)
	\rangle \right.
\nonumber \\
&+& [\bar{u}v]^{r}_{\alpha^{\prime}\beta^{\prime}} 
	(\mathbf{p}^{\prime}_1 ,\mathbf{p}^{\prime}_2 ,\mathbf{k};t) 
	\langle b_{\alpha}(-\mathbf{p}_1,t)
		a_{\beta}(\mathbf{p}_2 ,t) 
		a^+_{\alpha^{\prime}}(\mathbf{p}^{\prime}_1,t)
		b_{\beta^{\prime}}^+ (-\mathbf{p}^{\prime}_2 ,t)
	\rangle  
\nonumber \\
&+& [\bar{v}u]^{r}_{\alpha^{\prime}\beta^{\prime}} 
	(\mathbf{p}^{\prime}_1 ,\mathbf{p}^{\prime}_2 ,\mathbf{k};t) 
	\langle b_{\alpha}(-\mathbf{p}_1,t)
		a_{\beta}(\mathbf{p}_2 ,t) 
		b_{\alpha^{\prime}}(-\mathbf{p}^{\prime}_1,t) 
		a_{\beta^{\prime}}(\mathbf{p}^{\prime}_2 ,t)
	\rangle  
\nonumber \\
&+& \left.[\bar{v}v]^{r}_{\alpha^{\prime}\beta^{\prime}} 
	(\mathbf{p}^{\prime}_1 ,\mathbf{p}^{\prime}_2 ,\mathbf{k};t) 
	\langle b_{\alpha}(-\mathbf{p}_1,t) 
		a_{\beta}(\mathbf{p}_2 ,t) 
		b_{\alpha^{\prime}}(-\mathbf{p}^{\prime}_1,t) 
		b_{\beta^{\prime}}^+ (-\mathbf{p}^{\prime}_2 ,t)
	\rangle	\right\}~.  
\label{31e}
\end{eqnarray}
\end{widetext}

The remaining processes can be easily identified in Eqs.~(\ref{29e}) and (%
\ref{31e}), where the random phase approximation (RPA) is used as the
truncation procedure. That leads to the following decoupling rule for the
correlators of the type
\begin{widetext}
\begin{eqnarray}
\langle a_{\beta ^{\prime }}^{+}(\mathbf{p}^{\prime },t)
 a_{\beta }(\mathbf{p}_{2},t)A_{r^{\prime }}(\mathbf{k}^{\prime },t)
 A_{r}^{(+)}(\mathbf{k},t)\rangle
&\simeq &
\langle a_{\beta ^{\prime }}^{+}(\mathbf{p}^{\prime },t)
 a_{\beta }(\mathbf{p}_{2},t)\rangle
\langle A_{r^{\prime }}(\mathbf{k}^{\prime },t)
 A_{r}^{(+)}(\mathbf{k},t)\rangle~,  
\label{32e} \\
\langle b_{\alpha }(-\mathbf{p}_{1},t)a_{\beta }(\mathbf{p}_{2},t)
a_{\alpha^{\prime }}^{(+)}(\mathbf{p}_{1}^{\prime },t)
b_{\beta ^{\prime }}^{(+)}(-\mathbf{p}_{2}^{\prime },t)\rangle 
&\simeq &
\langle b_{\alpha }(-\mathbf{p},t)
        b_{\beta^{\prime }}^{(+)}(\mathbf{p}_{2}^{\prime },t)\rangle
\langle a_{\beta }(\mathbf{p}_{2},t)
  	a_{\alpha ^{\prime }}^{(+)}(\mathbf{p}_{1}^{\prime },t)\rangle~.
\label{32f}
\end{eqnarray}
\end{widetext}
As the result, only the first and the fourth term in Eq.~(\ref{29e}) and the
second term in Eq.~(\ref{31e}) survive in this order of
perturbation theory (Approximation 2).

The next approximation is the diagonalization of all one-particle
correlation functions with respect to the momentum variables and spin
indices,
\begin{eqnarray}
\langle a_{\alpha }^{+}(\mathbf{p},t)a_{\beta }(\mathbf{p}^{\prime
},t)\rangle &=&\delta _{\alpha \beta }\delta (\mathbf{p}-\mathbf{p^{\prime }}%
)f(\mathbf{p},t)~,  \label{6g} \\
\langle a_{\alpha }(\mathbf{p},t)a_{\beta }^{+}(\mathbf{p}^{\prime
},t)\rangle &=&\delta _{\alpha \beta }\delta (\mathbf{p}-\mathbf{p^{\prime }}%
)f^{c}(\mathbf{p},t)~,  \label{6'g}
\end{eqnarray}%
These relations mean that spin effects are neglected. 
The analogous approximation was introduced for the photon correlation 
function, Eq.~(\ref{25_2}).

The processes of the instantaneous radiation of two photons have been
omitted here, i.e., in Eq.~(\ref{29e}) the substitution $A_{r^{\prime}}(%
\mathbf{k}^{\prime},t)\rightarrow A_{r^{\prime}}^{(-)}(\mathbf{k}%
^{\prime},t) $ has been made (Approximation 3). In order to rewrite Eq.~(\ref%
{29e}) in the integral form, let us perform the intermediate transition to
the interaction representation
\begin{eqnarray}
\tilde{a}(\mathbf{p},t) &=& a(\mathbf{p},t)\exp\left\{i\int_{t_0}^{t}
dt^{\prime}\omega(\mathbf{p},t^{\prime})\right\},  \nonumber \\
\tilde{b}(\mathbf{p},t) &=& b(\mathbf{p},t)\exp\left\{i\int_{t_0}^{t}
dt^{\prime}\omega(\mathbf{p},t^{\prime})\right\},  \label{36e} \\
\tilde{A}^{(\pm)}(\mathbf{k},t) &=& A^{(\pm)}(\mathbf{k},t)\exp\left\{\mp
ik(t-t_0)\right\},  \nonumber
\end{eqnarray}
where $t_0$ is some initial time.

The approximations (\ref{32e})-(\ref{6'g}) allow to rewrite the anomalous
correlation functions from the l.h.s. of Eq.~(\ref{29e}) taking into account
the vacuum polarization contribution (\ref{31e}) so that
\begin{eqnarray}
&&\langle b_{\alpha }(\mathbf{-p}_{1},t)a_{\beta }(\mathbf{p}_{2},t)
  A_{r}^{(+)}(\mathbf{k},t)\rangle =
\nonumber\\
&&-\frac{ie\delta (\mathbf{p}_{2}-\mathbf{p}_{1}+\mathbf{k})}
{\sqrt{2k}(2\pi )^{3/2}}
\int^{t}dt^{\prime }[\bar{u}\upsilon ]_{\alpha \beta }^{r}
(\mathbf{p}_{2},\mathbf{p}_{1},\mathbf{k};t^{\prime })  
\nonumber \\
&&\times \big\{[f(\mathbf{p}_{1},t^{\prime })+f(\mathbf{p}_{2},t^{\prime
})-1][1+F_{r}(\mathbf{k},t^{\prime })]  \nonumber \\
&&+[1-f(\mathbf{p_{1}},t^{\prime })]
[1-f(\mathbf{p_{2}},t^{\prime })]\big\}
\mathrm{e}^{-i\theta (\mathbf{p}_{1},\mathbf{p}_{2},\mathbf{k};t^{\prime},t)}
\nonumber\\
&& +c.c.~,  
\label{7g}
\end{eqnarray}
where it was used that $f^{c}=1-f$ due to the electric charge neutrality of
the vacuum (\ref{1e}) at $t\rightarrow -\infty $ and
\begin{equation}
\theta (\mathbf{p}_{1},\mathbf{p}_{2},\mathbf{k};t^{\prime},t)
=\int_{t^{\prime }}^{t}d\tau \left[ \omega (\mathbf{p}_{1},\tau )
+\omega(\mathbf{p}_{2},\tau )-k\right] ~.  
\label{8g}
\end{equation}
The first group of terms in the curly brackets in Eq.~(\ref{7g}) corresponds
to the one-photon annihilation process (this contribution was investigated
in the work \cite{BCA}) while the second group describes the radiationless
vacuum fluctuations. In the case of a strong subcritical laser field the
number density of the radiated photons is small, $F_{r}(\mathbf{k},t)\ll
1$ (Approximation 4), so that the influence of the photon reservoir on the
photon emissivity of the system can be neglected. Eq.~(\ref{7g}) then takes
the form
\begin{eqnarray}
&&\langle b_{\alpha }(\mathbf{-p}_{1},t)
a_{\beta }(\mathbf{p}_{2},t)A_{r}^{(+)}(\mathbf{k},t)\rangle =
\nonumber\\
&&-\frac{ie\delta (\mathbf{p}_{2}-\mathbf{p}_{1}+\mathbf{k})}
{\sqrt{2k}(2\pi )^{3/2}} \int^{t}dt^{\prime }
\mathrm{e}^{-i\theta (\mathbf{p}_{1},\mathbf{p_{2}},\mathbf{k};t^{\prime},t)}  
\nonumber \\
&&\times \lbrack \bar{u}\upsilon ]_{\alpha \beta }^{r}
(\mathbf{p_{2}},\mathbf{p_{1}},\mathbf{k};t^{\prime })
f(\mathbf{p_{1}},t^{\prime })f(\mathbf{p_{2}},t^{\prime })
+c.c.~.  
\label{9g}
\end{eqnarray}
Substituting (\ref{9g}) into Eq.~(\ref{29e}), we obtain a closed expression
for the photon production rate
\begin{eqnarray}
\dot{F}(\mathbf{k},t)&=&
\frac{e^{2}}{4k(2\pi )^{3}}\int^{t}dt^{\prime }
\int d^{3}p \mathrm{e}^{-i\theta 
(\mathbf{p},\mathbf{p}+\mathbf{k},\mathbf{k};t^{\prime },t)}
\nonumber\\
&& K(\mathbf{p},\mathbf{p}+\mathbf{k},\mathbf{k};t,t^{\prime})
f(\mathbf{p},t^{\prime })f(\mathbf{p}+\mathbf{k},t^{\prime })
\nonumber\\
&& +c.c.~,
\label{10g}
\end{eqnarray}
where we have introduced the two-time convolution with respect to spin and
polarization indices
\begin{eqnarray}
K(\mathbf{p},\mathbf{p}+\mathbf{k},\mathbf{k};t,t^{\prime })
&=&[\bar{v}u]_{\beta \alpha }^{r}
(\mathbf{p},\mathbf{p}+\mathbf{k},\mathbf{k};t)~\nonumber\\
&&[\bar{u}\upsilon ]_{\alpha \beta }^{r}
(\mathbf{p}+\mathbf{k},\mathbf{p},\mathbf{k};t^{\prime })~.  
\label{11g}
\end{eqnarray}
with the definitions (\ref{5e}) and (\ref{21e}).

Additionally, it is assumed in Eq.~(\ref{10g}) that the photons have
equiprobable distributions regarding their polarizations, $F_1=F_2=F$.

Thus, the photon production rate is a nonlinear (quadratic) non-Markovian
function with respect to the electron-positron distribution function $f(%
\mathbf{p},t)$. This order in the nonlinearity corresponds to the result
\cite{Rit} of the S-matrix approach.

The consequent estimation procedure of the integrals on the l.h.s. of 
Eq.~(\ref{10g}) (method of photon count) was presented in \cite{BCA} on the
basis of the methods given in \cite{BIP}. The meaning of these
approximations is the following. On the r.h.s. of Eq.~(\ref{10g}) there is a
high frequency multiplier $\exp(-i\theta)$ with the phase (\ref{8g}). In
order to select the low frequency component of the photon production rate (%
\ref{10g}) (this corresponds to the averaged, observable value), it is
necessary to compensate this high frequency phase by means of the higher
harmonics in the Fourier decompositions of the other functions in the
integral (\ref{10g}). Indeed, the integrand in the expression (\ref{8g}) for
the phase (mismatch) is very large
\begin{equation}
\omega(\mathbf{p},\tau)+\omega(\mathbf{p}+\mathbf{k},\tau)-k\sim 2m
\label{50e}
\end{equation}
and the energy conservation law is not fulfilled for the one-photon
annihilation process. That is why the radiation of a real photon can be
interpreted as a multiphoton process. The photon number $N_{\nu}$ from the
photon condensate of the external quasiclassical {laser field} with
the frequency $\nu$ can be estimated from the condition to compensate the
mismatch (\ref{50e}) by the energy of $N_{\nu}$ quasiclassical photons. This
leads to {the estimate} $N_{\nu}\sim 2m/\nu$. For optical lasers
this is a huge number and therefore such kind of fluctuation event is very
scarce. But in the case of $\gamma-$ radiation the result raises hopes for
the possibility of an observation. In the general case, the effect is
defined by the external field parameters: the amplitude $E_0$ and the
frequency $\nu$.

This conclusion about the role of multiphoton processes correlates with the
analysis of the absorption coefficient of the electron - positron plasma
created from vacuum in the infrared region \cite{BIP}. 
According to the structure of the time dependent $u,\upsilon$-spinors 
\cite{PNN}, the convolution (\ref{11g}) is a polynomial in $e A(t)$
and hence it can not guarantee for the necessary compensation. 
Below it is assumed that the laser electric field 
$A(t)=A^{3}(t)=-(E_0/\nu) \cos(\nu t) $ is {subcritical}, $E_0\lesssim E_c$.
Therefore, we use here the Markovian approximation (Approximation 5) 
$K(\mathbf{p},\mathbf{p}+\mathbf{k},\mathbf{k};t,t^{\prime})
\rightarrow K(\mathbf{p},\mathbf{p}+\mathbf{k},\mathbf{k};t,t)= K_0= 4$, 
see Appendix A. 
Thus there are two sources for compensation of the high-frequency phase 
(\ref{8g}): the time dependence of the fermion distribution functions 
$f(\mathbf{p},t)$ and the multiphoton contributions contained in the mismatch 
(\ref{50e}).

As it can be seen from the KE's (\ref{KE-8}), (\ref{KE-9}) the fermion
distribution function $f(\mathbf{p},t)$ has two time scales defined by the
frequency $\nu$ of the external field and by the rest mass $m$ (or the
one-particle quasi-energy $\omega(\mathbf{p},t)$). Then the corresponding
Fourier transformation can be defined by the double series
\begin{equation}
f(\mathbf{p},t)=\sum_{n,l}f_{n,l}(\mathbf{p})e^{in\nu t+imlt}.  \label{57e}
\end{equation}
The numerical calculations have shown \cite{PRL} that the low-frequency
behavior of the fermion distribution function in the presence of a
linearly polarized field is defined approximately by the second harmonic
(Approximation 6)
\begin{equation}
f(\mathbf{p},t)=\bar{f}(\mathbf{p})\left[1-\cos(2\nu t) \right]/2~.
\label{second}
\end{equation}
The higher low-frequency harmonics and the high-frequency harmonics in the
decomposition (\ref{57e}) are very small compared to (\ref{second}) and will
be omitted in the following.

The choice of the following approximations depends on the value of the
adiabaticity parameter \cite{Popov}
\begin{equation}
\gamma = (E_c/E_0)/(\nu/m)~.  \label{gamma}
\end{equation}
The domain $\gamma\gg 1$ corresponds to the multiphoton processes in a
rather small external field $E_0\ll E_c$. In the case $\gamma \ll 1$ the
external field is rather large, $E_0\lesssim E_c$. These two limiting cases
will be considered below.

\subsection{Multiphoton domain ($\protect\gamma\gg 1$)}

Since $E_{0}\ll E_{c}$, one can use the appropriate perturbation
theory in order to select the constant component in the photon
production rate. The
unique source is now the time dependence of the mismatch (\ref{50e}). 
For $E_{0}\ll E_{c}$, one can select in the quasi-energy the first order field
effect (Approximation 7) 
\begin{equation}
\omega (\mathbf{p},t)\simeq \omega _{0}(\mathbf{p})
-\frac{e\mathbf{E}_{0}\mathbf{p}}{\nu \,\omega _{0}(\mathbf{p})}\cos (\nu t)~,  
\label{51e}
\end{equation}
where $\omega _{0}(\mathbf{p})=\omega (\mathbf{p},t)|_{A=0}$. 
Then the phase (\ref{8g}) will be
\begin{equation}
\theta (t,t^{\prime })\simeq \Omega _{0}(\mathbf{p},\mathbf{k})(t-t^{\prime
})+a(\mathbf{p},\mathbf{k})[\sin \nu t-\sin \nu t^{\prime }]~,  \label{52e}
\end{equation}%
where the mismatch in the absence of the external field is
\begin{equation}
\Omega _{0}(\mathbf{p},\mathbf{k})=
\omega _{0}(\mathbf{p})+\omega _{0}(\mathbf{p}+\mathbf{k})-k~,  
\label{53e}
\end{equation}
and
\begin{equation}
a(\mathbf{p},\mathbf{k})=
-\frac{e}{\nu ^{2}}\left[\frac{\mathbf{E}_{0}\mathbf{p}}{\omega _{0}(\mathbf{p})}
+\frac{\mathbf{E}_0(\mathbf{p}+\mathbf{k})}{\omega_0(\mathbf{p}+\mathbf{k})}
\right] ~.  
\label{54e}
\end{equation}
The fast oscillating function in Eq.~(\ref{10g}) allows then the
following representation
\begin{equation}
\cos \Phi (t,t^{\prime })\simeq e^{i\Omega _{0}(t-t^{\prime
})}\sum_{n,n^{\prime }}J_{n}(a)J_{n^{\prime }}(a)e^{i\nu (nt-n^{\prime
}t^{\prime })}+c.c.~,  \label{55e}
\end{equation}
which is based on the known decomposition
\begin{equation}
\exp (ia\sin \varphi )=\sum_{n=-\infty }^{\infty }J_{n}(a)e^{in\varphi },
\label{56e}
\end{equation}
where $J_{n}(a)$ is the Bessel function. 
Thus, the representation (\ref{55e}) contains two high frequency harmonics 
$\omega _{0}(\mathbf{p})$ and $\omega _{0}(\mathbf{p}+\mathbf{k})$ 
(see Eq.~(\ref{53e})), the set of the low frequency harmonics $n\nu $ and the 
$k$- harmonic corresponding to the radiated photon. 

The following steps are these: we calculate the time integral in 
Eq.~(\ref{10g}) and select its time independent component corresponding to
the observed value. Performing then the remaining momentum space
integrations in the isotropic approximation and using the textbook formula
\begin{equation}
\delta \lbrack \phi (x)]=\sum_{i}\left\{ |\phi ^{\prime }(x_{i})|\right\}
^{-1}\delta (x-x_{i}),\quad \phi (x_{i})=0,  
\label{62e}
\end{equation}
we obtain then in the low frequency approximation 
\begin{eqnarray}
\dot{F}(k)&=&\frac{\alpha K_{0}}{2k}\frac{p_{1}\omega _{1}\sqrt{\omega
_{1}^{2}+k^{2}}}{\omega _{1}+\sqrt{\omega _{1}^{2}+k^{2}}}J_{n_{0}+1}(a)
\nonumber\\
&&\left[ J_{n_{0}+3}(a)+J_{n_{0}-1}(a)\right] \bar{f}(p_{1})\bar{f}(p_{1}+k)~,
\label{12g}
\end{eqnarray}
where we took into account the lowest harmonics of the distribution function
Eq.~(\ref{second}) and $\omega _{1}=\sqrt{m^{2}+p_{1}^{2}}$. 
The argument of the Bessel function in the isotropic approximation is
\begin{equation}
a=\frac{2\sqrt{\pi \alpha }E_{0}}{\nu ^{2}}\left[ \frac{p_{1}}{\omega _{1}}
+ \frac{p_{1}+k}{\sqrt{\omega _{1}^{2}+k^{2}}}\right] ~.  
\label{13g}
\end{equation}
Let us explain the meaning of the momentum $p_{1}$. 
By $p_{0}$ we denote the positive root of the equation $\Omega _{0}-n\nu =0$,
where
\begin{equation}
\Omega _{0}(p,k)=\omega _{0}(p)+\sqrt{\omega _{0}^{2}(p)+k^{2}}-k
\label{14g}
\end{equation}
is the mismatch and $n=\left[ \Omega _{0}(p_{0},k)/\nu \right] $ ($[x]$ is
the integer part $x$) is the photon number necessary for overcoming the
mismatch (\ref{14g}). Then we obtain
\begin{equation}
p_{0}=\left\{\frac{(n\nu)^2(n\nu -2k)^2}{4(n\nu -k)^2}-m^2\right\}^{1/2}~.  
\label{15g}
\end{equation}
Let now $n_{0}$ be the minimal number of quasiclassical photons, 
$n_{0}\sim p_{0}=0$. 
The photon production rate (\ref{15g}) is equal to zero in this point. 
Let us suppose $n_{1}=n_{0}+1$ in Eq.~(\ref{15g}). 
The momentum $p_{1}$ corresponds to this number, $p_{1}=n_{1}$. 
Thus, according to Eq. (\ref{16g}) we have 
$I(k)\sim \alpha ^{n_{0}+1}\bar{f}^{2}(0)$. 
It is difficult to estimate the general order of the perturbation theory, 
because the amplitude $\bar{f}(0)$ of the distribution function is
calculated nonperturbatively.

\subsubsection{The case of optical vacuum excitation ($\nu\ll m$)}

In the optical part of the photon spectrum ($k\lesssim \nu $) we have 
$p_{1}$=$\sqrt{km}$ as the first root of the equation 
$\Omega_{0}(p_{1})-(n_{0}+1)\nu =0$ 
(it corresponds to the leading contribution from the set $n>n_{0}$), 
$a=2(E_{0}/E_{c})(m/\nu )^{3/2}\ll 1$ and $n_{0}=[2m/\nu ]\gg 1$, i.e., 
the necessary number of quasiclassical photons
is huge and the intensity of photon radiation a very small.
Indeed, from Eq.~(\ref{12g}) it follows that
\begin{eqnarray}
I(k)&=&\frac{1}{m}\frac{dF(k)}{dt}\nonumber\\
&=&\frac{\alpha}{4} K_{0}\sqrt{\frac{m}{k}}\bar{f}^{2}(0)
\left\{\frac{1}{n_{0}^{2}}\left[\frac{E_{0}}{E_{c}}
\left(\frac{m}{\nu }\right)^{3/2}\right]^{2n_{0}}\right\} ~.  
\label{16g}
\end{eqnarray}
The value $\bar{f}(0)$ can be estimated as a result of the numerical
solution of the KE (\ref{KE-8}), (\ref{KE-9}) for $e^{-}e^{+}$ excitations
in a laser field \cite{Smolyansky:2010as}. 
For the PW laser system Astra Gemini we have $E_{0}\sim 10^{-5}E_{c}$, 
$\lambda =800$ nm and $\bar{f}_{2}(0)\sim 10^{-11}$ \cite{PRL,Blaschke:2008wf}.
The spectral distribution of the radiated photons from the volume 
$\lambda ^{3}$ of the focal spot per time interval,
\begin{equation}
\frac{dN_{k}}{dtdk}=
\frac{\lambda ^{3}}{\pi ^{2}}I(k)~k^{2}m=\frac{8\pi k^{2}m}{\nu ^{3}}I(k)~,  
\label{17g}
\end{equation}
will be negligibly small for the mentioned parameters. 
However, the term in the curly brackets on the r.h.s. of Eq.~(\ref{16g}) 
behaves as a $\theta (a)$-function with the branch point $a_{0}=2$ when 
letting $a\rightarrow a_{0}$.
Then the spectral distribution (\ref{17g}) starts to grow strongly.
Unfortunately, this value $a_{0}=2$ lies outside of the validity range of
Eq.~(\ref{16g}). 
Nevertheless, this gives a hint on the possible growth of the radiation 
intensity in this domain. 
Some additional analyses are necessary here.

The $\gamma$-ray part of the photon spectrum $k \sim m$ can not be
considered in the framework of this approximation ($a\gg 1$ again).

\subsubsection{The case of $\gamma$-ray vacuum excitation ($\nu\sim m$)}

In this case the mismatch (\ref{14g}) can be compensated by the smallness of
the photon number from the quasiclassical laser field, $n_0\gtrsim 1$
\cite{Smolyansky:2010as}. 
Let $n_0=1$ (this is the hypothetical limiting case, e.g., for the planned XFEL
facility with $\lambda=0.15$ nm \cite{Ringwald}). 
The developed theory is working well in this case ($a\ll 1 $) for the 
subcritical fields $E\ll E_c$.

In the optical part of the photon band ($k\ll m$) we have $p_1=\sqrt{3}m$ and
\begin{equation}
a=\sqrt{3}\frac{E_0}{E_c}\left(\frac{m}{\nu} \right)^2~.  \label{18g}
\end{equation}
The spectral distribution according to Eq.~(\ref{12g}) is
\begin{equation}
\frac{dN_k}{dtdk}=\frac{3\sqrt{3}\pi \alpha K_0 k}{2 \nu}\bar{f}^2(p_1)
\left(\frac{E_0}{E_c} \right)^2\left(\frac{m}{\nu} \right)^6~.  
\label{19g}
\end{equation}
Thus, the effect grows linearly with $k$. 
For $\nu=1$ MeV, $E_0=10^{-5}E_c$ and $\bar{f}(p_1)\sim 10^{-11}$ 
(see Fig.~\ref{fig:fig1}) we obtain again a negligibly small effect:
the suppression factor is $\bar{f}(0)E_0/E_c \sim 10^{-5}$ so that a very
weak signal results.

The situation is slowly changing when going to higher frequencies of the
excited signal ($X$-ray or $\gamma $-ray domain) at $E/E_{c}=$ const. 
One can demonstrate this by writing $p_{1}$ (\ref{15g}) for $k\neq 0$ and 
$n_{0}=1$
\begin{equation}
p_{1}=m\left\{ 4\left[ \frac{2+k/m}{2+k/2m}\right] ^{2}-1\right\} ^{1/2}~.
\label{20g}
\end{equation}

The situation becomes more optimistic at $E\rightarrow E_c$ when $a\to 1$.
For example, for the XFEL with $E=0.24~E_c$ and $\lambda=0.15$ nm 
\cite{Ringwald} the intensity (\ref{19g}) can be accessible to observation,
apparently. 
However, this case needs special investigation since for $\gamma\lesssim 1$ 
the presented approach is not valid.

\subsection{Strong field case ($\protect\gamma \lesssim 1$)}

We will use here the effective mass model \cite{Rit} based on the
approximation
\begin{eqnarray}
\omega (\mathbf{p},t) &=&
\sqrt{m^{2}+\left( \mathbf{p}-e\mathbf{A}(t)\right)^{2}}\rightarrow  
\omega _{\ast }(p)~, \nonumber \\
\omega _{\ast }(p) &=&\sqrt{m_{\ast }^{2}+\mathbf{p}^{2}},
\label{3h}
\end{eqnarray}
with the effective mass defined by the relation
\begin{eqnarray}
m_{\ast }^{2} &=&m^{2}+e^{2}\prec \mathbf{A}^{2}(t)\succ
=m^{2}+e^{2}E_{0}^{2}/2\nu ^{2}  \nonumber \\
&=&m^{2}(1+1/2\gamma ^{2}),  \label{4h}
\end{eqnarray}
where $\prec ...\succ $ denotes the time averaging operation,
and $\gamma $ is the adiabaticity parameter (\ref{gamma}).

In this approximation the phase (\ref{8g}) becomes monochromatic
\begin{equation}
\theta (\mathbf{p}_{1},\mathbf{p}_{2},k;t^{\prime },t)
=\Omega _{\ast }(\mathbf{p}_{1},\mathbf{p}_{2},\mathbf{k})(t-t^{\prime }),  
\label{5h}
\end{equation}
\begin{equation}
\Omega _{\ast }(\mathbf{p}_{1},\mathbf{p}_{2},\mathbf{k})
=\omega _{\ast }(\mathbf{p}_{1})+\omega _{\ast }(\mathbf{p}_{2})-k,  
\label{6h}
\end{equation}
i.e., the approximation (\ref{3h}) leads to a suppression of multiphoton
processes (it corresponds to large values of the adiabacity parameter 
$\gamma \gg 1$) and the mismatch (\ref{6h}) can be compensated by the
harmonics of the fermion distribution functions in Eq.~(\ref{10g}) only.

The inspection of the fermion distribution function shows, in particular,
that it oscillates basically with twice the laser frequency (\ref{second}).
The substitution of Eqs.~(\ref{5h}) and (\ref{second}) into the KE (\ref{10g}) 
allows to perform the time integration, leading to the appearance of two
harmonics in the radiation spectrum only (the 2$^{\mathrm{nd}}$ and the 
4$^{\mathrm{th}}$),
\begin{equation}
\dot{F}(\mathbf{k},t)=-A^{(2)}(\mathbf{k})
\cos (2\nu t)+A^{(4)}(\mathbf{k})\cos (4\nu t),  \label{8h}
\end{equation}
\begin{equation}
A^{(2)}(\mathbf{k})=\frac{\pi ^{2}K_{0}\alpha }{2k}\int \frac{d^{3}\mathbf{p}
}{(2\pi )^{3}}\bar{f}(\mathbf{p})\bar{f}(\mathbf{p}+\mathbf{k})\delta \left(
2\nu -\Omega _{\ast }\right) ~,  \label{9h}
\end{equation}
\begin{equation}
A^{(4)}(\mathbf{k})=\frac{\pi ^{2}K_{0}\alpha }{8k}\int \frac{d^{3}\mathbf{p}
}{(2\pi )^{3}}\bar{f}(\mathbf{p})\bar{f}(\mathbf{p}+\mathbf{k})\delta \left(
4\nu -\Omega _{\ast }\right) ~.  \label{10h}
\end{equation}

It is important that a constant component is absent here, because the
mismatch (\ref{50e}) could not be compensated in this case by other sources
of the time dependence on the r.h.s. of Eq.~(\ref{10g}).\footnote{
This is in contrast to the case $\gamma \gg 1$, where accounting for
multi-photon processes in the phase (\ref{8g}) leads to a constant component
\cite{Smolyansky:2010as}.}

Thus, in the case of the infinite system the solution (\ref{8h}) can be
interpreted as "breathing" of the photon subsystem. 
However, the situation is changed, when the generation of the 
$e^{-}e^{+}\gamma $ plasma is considered in a small spatial domain of the 
focal spot with volume $\sim \lambda ^{3}$ due to the vacuum condition of the 
absence of the $e^{-}e^{+}\gamma $ plasma in the initial moment of switching 
on the laser field. 
In this case one can expect, that all annihilation photons generated
in the first half-period of the field will leave the volume of the system
and therefore in the next half-period the reverse process (photon
transformation to $e^{-}e^{+}$ plasma) will be impossible. 
Such a mechanism leads to a pulsation pattern for the photon radiation from 
the focal spot.
It corresponds to introducing the condition of a positive definite photon
production rate on the r.h.s. of Eq.~(\ref{8h}).

For estimates of the amplitudes (\ref{9h}), (\ref{10h}) let us introduce
the additional model approximation in the spirit of the model (\ref{3h}),
\begin{equation}
\omega_\ast(\mathbf{p}+\mathbf{k})\rightarrow \omega_\ast(p,k)
=\sqrt{\omega_\ast^{2}(p)+k^{2}}~,  
\label{11h}
\end{equation}
and the isotropisation condition 
$\bar{f}(\mathbf{p}+\mathbf{k})\rightarrow \bar{f}(p+k)$. 
The integrals on the r.h.s of Eqs.~(\ref{9h}), (\ref{10h}) can then be 
calculated. For example,
\begin{equation}
A^{(2)}(\mathbf{k})
=\frac{\alpha}{k}\bar{f}(p_{0})\bar{f}(p_{0}+k)
\frac{\omega_\ast(p_{0})\omega _{\ast }(p_{0},k)}{\omega_\ast(p_{0})
+\omega_\ast(p_{0},k)}p_{0},  
\label{12h}
\end{equation}
where
\begin{equation}
p_{0}=\sqrt{\frac{4\nu ^{2}(k+\nu )^{2}}{(k+2\nu )^{2}}-m_{\ast }^{2}}
\label{13h}
\end{equation}
is the root of the equation $\Omega _{\ast }-2\nu =0$. From Eq. (\ref{13h})
follows the threshold condition\footnote{%
A similar effect was found first in the theory describing the absorption of
a weak signal by the $e^{-}e^{+}$ plasma created from vacuum \cite{BIP}.}
\begin{equation}
\frac{2\nu (k+\nu )}{k+2\nu }\geqslant m_{\ast }~.  
\label{14h}
\end{equation}
This condition is rather nontrivial because the effective mass (\ref{4h})
depends also on $\nu $. The minimal permissible value $\nu =2m_{\ast }$
corresponds to $k=0$. 
For the 4$^{\mathrm{th}}$ harmonic the threshold value falls to 
$\nu =m_{\ast }$, which is close to the parameters of the XFEL 
\cite{Ringwald}.

The $1/k$ - dependence on the r.h.s. of Eq.~(\ref{12h}) corresponds to the
flicker-like noise of electrodynamical origin. This feature in the spectrum
of radiated annihilation photons has been found first in 
Ref.~\cite{Blaschke:2010vs}.

The number of photons with the frequency $k$ lying in the interval $[k,k+dk]$
and radiated from the focal spot with the volume $\lambda ^{3}=\nu ^{-3}$
per time interval is defined by Eq.~(\ref{17g}). The fraction on the r.h.s.
of Eq.~(\ref{12h}) is a slow function of the frequencies $k$ and $\nu $ and
for the sake of a preliminary estimate it can be replaced by $m_{\ast }/2$.
For the 2$^{\mathrm{nd}}$ harmonic we then obtain from Eqs.~(\ref{12h}) and 
(\ref{17g})
\begin{equation}
\frac{d^{2}N^{(2)}}{dtdk}=
\frac{2\pi \alpha K_{0}km_{\ast }}{\nu ^{3}}\bar{f}(p_{0})\bar{f}(p_{0}+k)p_{0}~. 
\label{16h}
\end{equation}

As a representative characteristics of the effectiveness of the radiation
from the focal spot domain we will consider the total photon number per time
interval,
\begin{equation}
\dot{N}^{(2)}=\frac{2\pi \alpha K_0 m_\ast}{\nu^3}
\int\limits_{0}^{k_{\max }}dk~k~\bar{f}(p_{0})\bar{f}(p_{0}+k)p_{0}~.
\label{17h}
\end{equation}
The electron and positron distribution functions entering here are
defined as the solutions of the corresponding non-perturbative
KE (Sect.~II) describing vacuum creation of $e^{-}e^{+}$
pairs under the action of a strong, time dependent electric field of
a standing wave of two counter-propagating laser beams. The cutoff
parameter $k_{\max }=2m_{\ast }$ is introduced in order to take into
account the annihilation photons in the radiation spectrum.

The fermion distribution function $f(\mathbf{p},t)$ is a rapidly decreasing
function with its maximum in the point $\mathbf{p}=0$ 
\cite{PRL,Blaschke:2008wf}. 
On this basis for a rough estimate one can put $p_{0}=0$ in the
arguments of these functions on the r.h.s. of Eq.~(\ref{17h}),
\begin{equation}
\dot{N}^{(2)}=\frac{2\pi \alpha K_{0}m_{\ast }}{\nu ^{3}}\bar{f}%
(0)\int\limits_{0}^{k_{\max }}dk~k~\bar{f}(k)p_{0}~,  \label{18h}
\end{equation}%
where according to Eq.~(\ref{13h})
\begin{eqnarray}
p_0(k) &=&\frac{m_\ast^2}{k+4m_\ast}\sqrt{48+56\frac{k}{m_\ast}
+15\frac{k^2}{m_\ast^2}} 
\nonumber \\
&\simeq &\frac{m}{4}\sqrt{48+56\frac{k}{m_{\ast }}}~,  
\label{19}
\end{eqnarray}
because the small $k_{\max} \ll m_\ast$ is effective in the integral (\ref{18h}).
As the result, we obtain the following order of magnitude estimate
\begin{equation}
\dot{N}^{(2)}\sim \alpha m_{\ast }\bar{f}^{2}(0)~.  
\label{20h}
\end{equation}
\begin{figure}
\includegraphics[width=0.47\textwidth]{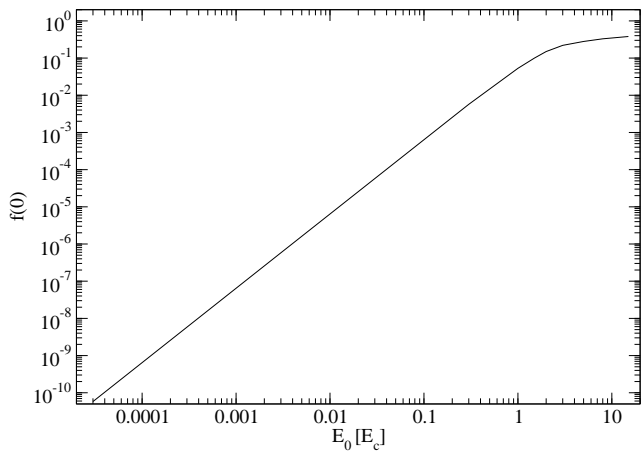}
\caption{Amplitude of the photon distribution function $\bar{f}(0)$ as a 
function of the dimensionless laser field strength $E_0/E_c$.}
\label{fig:fig1}
\end{figure}
For the XFEL\ parameters $E_{0}=0.24~E_{c}$ and $\lambda =15$ nm 
\cite{Ringwald} we have according to the kinetic theory in the $e^{-}e^{+}$
sector $\bar{f}(0)\sim 10^{-2}$, see Fig.~\ref{fig:fig1}. 
From Eq.~(\ref{20h}) then follows
\begin{equation}
\dot{N}^{(2)}\sim 10^{17}~~\mathrm{s}^{-1}~.  
\label{21h}
\end{equation}
For the 4$^{\mathrm{th}}$ harmonic with the oscillation amplitude (\ref{10h}) 
the threshold for the generation of annihilation photons is lowered (see
discussion after Eq.~(\ref{14h})) but the intensity of the photon radiation
is also lowered so that the order of magnitude of (\ref{21h}) remains
unchanged.

\section{Summary}

We have studied the photon production rate resulting from the one-photon
annihilation mechanism in a quasiparticle EPP created from the vacuum under
the influence of a strong electric field being a necessary condition for the
possible occurrence of such a process \cite{Rit}. The required strong
electric fields occur, e.g., in the focal spot of two counter-propagating
high-intensity laser beams. The methodic basis is an appropriately developed
kinetic theory constructed in the quasiparticle representation. The fermion
sector of the theory has been investigated before \cite%
{Schmidt:1998vi,Blaschke:2008wf} on an essentially nonperturbative basis.
The photon kinetics can be considered in the framework of the usual
perturbation theory which allows to truncate the BBGKY hierarchy of KEs at
the lowest order with respect to the fine-structure constant and to obtain
the closed formula (\ref{10g}) for the photon production rate including
vacuum polarization effects.

We have investigated this expression for different characteristics
of the laser field (frequencies $\nu$ and amplitudes $E_0$)
which is conveniently described with the adiabaticity parameter
$\gamma$. In order to become observable it is necessary for the
photon radiation process to compensate the energy mismatch
(\ref{50e}) by a sufficient number of quasiclassical photons from
photon reservoir of the external electric field. The strength of the
laser field $E_0$ defines the intensity of this process. Thus,
it can be understood that in the case of optical lasers the number
of quasiclassical photons in the optical range needed for the
compensation of the mismatch (\ref{50e}) is huge and the intensity
of such process is negligibly small.

However, the intensity of photon production increases strongly in the domain
of $\gamma-$ray excitations of the vacuum (where only a small number of
quasiclassical photons is necessary to overcome the mismatch) and reaches
the values sufficient for experimental observation, e.g., for the projected
XFEL \cite{Ringwald}. 
This result is qualitatively confirmed by the recent findings of 
Ref.~\cite{Ilderton:2011ja} which have been obtained within a different 
approach.

In the case of the two-photon annihilation channel the necessity to
compensate the mismatch of the type (\ref{50e}) is absent
and therefore the photon production rate can be large enough to be
observable already in the optical domain. This channel of the EPP
annihilation will be investigated separately.

We would like to note also that the type of kinetic theory based on the
quasiparticle representation, which has been described in the present work
can also be useful for the investigation of other reaction channels in the 
e$^-$e$^+\gamma$ plasma. 
One of such channels is the cascade process of EPP multiplication in a strong 
electric field (e.g., \cite{Elkina:2010up}). 
The example of one-photon annihilation presented here shows, that strong 
external fields can lead to qualitative modifications of processes relative 
to standard QED.

\section*{Acknowledgements}

We thank A.~V.~Tarakanov for his contributions at an early stage of this
work and A.~G.~Lavkin for his help in solving the KE (17) as the basis
for Fig.~1. 
Special thanks go to S.~M.~Schmidt for his long-standing collaboration with 
us in developing the kinetic approach to pair production in external fields.
D.B. and S.A.S. acknowledge support during their visit at the
Helmholtz Forschungszentrum J\"ulich, where this work has been started.
S.A.S. is grateful for hospitality and support at the Institute for
Theoretical Physics of the University of Wroclaw and the Bogoliubov Laboratory
for Theoretical Physics of the JINR Dubna, where this work was completed.



\begin{appendix}
\section{Convolution (\ref{11g})}

As an example, we will consider here the calculation of the convolution
(\ref{11g}) in the Markovian approximation.
Setting $t=t^{\prime }$, we obtain
\begin{eqnarray}
&&K(\mathbf{p}_{1},\mathbf{p}_{2},\mathbf{k};t,t)=\nonumber\\
&&=[\bar{v}_{\alpha }(\mathbf{p}_{1})\gamma ^{\mu }u_{\beta}(\mathbf{p}_{2})]
[\bar{u}_{\beta } (\mathbf{p}_{2})\gamma ^{\mu}v_{\alpha }(\mathbf{p}_{1})]
e_{\mu }^{r} (\mathbf{k})e_{\nu}^{r}(\mathbf{k})\ . 
\nonumber\\
\label{A1}
\end{eqnarray}
With help of the relations \cite{DeGroot:1980dk}
\begin{eqnarray}
\sum_{\alpha }[u_{\alpha }(\mathbf{p})]_{k}[\bar{u}_{\alpha }(\mathbf{p})]_{i}
&=&\frac{1}{2\omega (p)}[\hat{p}+m]_{ki},  \nonumber \\
\sum_{\alpha }[v_{\alpha }(\mathbf{p})]_{k}[\bar{v}_{\alpha }(\mathbf{p})]_{i}
&=&\frac{1}{2\omega (p)}[\hat{p}-m]_{ki}
\label{A2}
\end{eqnarray}
one can transform Eq.~(\ref{A1}) to the form
\begin{eqnarray}
&&K(\mathbf{p}_{1},\mathbf{p}_{2},\mathbf{k};t,t) =
\nonumber\\
&&= \frac{1}{4\omega (p_{1})\omega (p_{2})}{\rm Tr}(\hat{p}_{1}-m) \gamma ^{\mu
}(\hat{p}_{2}+m)\gamma ^{\nu }e_{\mu }^{r}(k)e_{\nu }^{r}(k)
\nonumber \\
&&=\frac{1}{\omega (p_{1})\omega (p_{2})}\left\{
2(p_{1}e^{r})(p_{2}e^{r})-\left( m^{2}+p_{1}p_{2}\right) \left(
e_{\mu}^{r}e_{\nu }^{r}g^{\mu \nu }\right) \right\}~. \nonumber\\
\label{A3}
\end{eqnarray}
The photon polarization vector $e_{\mu }^{r}(\mathbf{k})$ is satisfying to
the conditions ( $r=1,2$)
\begin{eqnarray}
g^{\mu \nu }e_{\mu }^{r}e_{\nu }^{r} &=&-2,
\nonumber \\
(p_{1}e^{r})(p_{2}e^{r})
&=&(\mathbf{p}_{1}\mathbf{e}^{r})(\mathbf{p}_{2} \mathbf{e}^{r})~.
\label{A4}
\end{eqnarray}
Thus, we finally obtain
\begin{eqnarray}
&&K(\mathbf{p}_{1},\mathbf{p}_{2},\mathbf{k};t,t) =
\nonumber\\
&&=\frac{2}{\omega(p_{1})\omega (p_{2})} \left\{
m^{2}+p_{1}p_{2}+(\mathbf{p}_{1}\mathbf{e}^{r})
(\mathbf{p}_{2}\mathbf{e}^{r})\right\}. 
\label{A5}
\end{eqnarray}

In the case $\mathbf{p}_{1}=\mathbf{p}_{2}=0$ it follows from (\ref{A5})
\begin{equation}
K(\mathbf{p}_{1},\mathbf{p}_{2},\mathbf{k};t,t)=K_{0}=4~. 
\label{A6}
\end{equation}

The last formula is used in the qualitative estimates of the photon
production rate in the different models.

For $\mathbf{p}_{1}\neq\mathbf{p}_{2}\neq0$ one can obtain another estimate.
Since in the integral we have
\begin{equation*}
p_{1}^{i}p_{2}^{k}\Rightarrow\frac{1}{3}\delta
_{ik}\mathbf{p}_{1}\mathbf{p}_{2},
\end{equation*}
we obtain from Eq.~(\ref{A6})
\begin{eqnarray}
&&K(\mathbf{p}_{1},\mathbf{p}_{2},\mathbf{k};t,t)=
\nonumber\\
&&=\frac{2}{\omega(p_{1})\omega (p_{2})}\left\{ m^{2}
+\omega (p_{1})\omega (p_{2})-\frac{1}{3}\mathbf{p}_{1}\mathbf{p}_{2}\right\}.
\label{A7}
\end{eqnarray}
\end{appendix}

\end{document}